\documentclass[twocolumn,american,prl]{revtex4-1}
\usepackage{lmodern}
\usepackage[T1]{fontenc}
\usepackage[latin9]{inputenc}
\usepackage{amsmath}
\usepackage{amssymb}
\usepackage{graphicx}

\makeatletter

\usepackage{color}

\makeatother

\usepackage{babel}
\begin{document}

\title{Second order functional renormalization group approach to one-dimensional systems in real and momentum space}

\author{Björn Sbierski and Christoph Karrasch}

\affiliation{Dahlem Center for Complex Quantum Systems and Institut für Theoretische
Physik, Freie Universität Berlin, 14195, Berlin, Germany}

\date{\today}
\begin{abstract}
We devise a functional renormalization group treatment for a chain
of interacting spinless fermions which is correct up to second
order in the interaction strength. We treat both inhomogeneous systems
in real-space as well as the translational invariant case in a k-space
formalism. The strengths and shortcomings of the different schemes as well as technical details of their implementation are discussed. We use the method to study two proof-of-principle problems in the realm of Luttinger liquid physics,
namely reflection at interfaces and power laws in the occupation number
as a function of crystal momentum. 
\end{abstract}
\maketitle

\section{Introduction}

The functional renormalization group (fRG) is a specific realization
of the RG concept on the basis of vertex functions for many-particle
systems \cite{Metzner2012,Kopietz2010}. The idea is to introduce an infrared cutoff
$\Lambda$ in the bare Green function of the system at hand, $G_{0}\rightarrow G_{0}^{\Lambda}$
with the limits $G_{0}^{\Lambda=\infty}=0$ and $G_{0}^{\Lambda=0}=G_{0}$.
Beyond these requirements, the nature of this cutoff is generic and
a particular choice will be presented below. Naturally, all vertices now depend on $\Lambda$; the ensuing fRG flow
equations quantify the change of the vertex functions
with $\Lambda$ and represent a coupled set of differential equations
organized in an infinite hierarchy of Feynman diagrams. In the limit
$\Lambda=\infty$, by virtue of the trivial dynamics, the vertex functions
are known exactly. Integration of the flow equations from $\Lambda=\infty$
to $\Lambda=0$ in principle yields an exact expression for the vertex functions of the
underlying problem. In practice, the infinite hierarchy of
flow equations has to be truncated; fRG results are then only an approximation, which, however, goes beyond perturbation theory since an infinite re-summation
of Feynman diagrams is included. Further, the approximations can ensure
the vertex functions to be exact up to some order $n$ in the interaction
strength, in this case we call this formalism fRG-$n$.

While the fact that only certain types of Feynman diagrams are included in the fRG methodology is a very delicate point
that requires a careful assessment of its approximate predictions on a case-by-case
basis, the approach offers a range of benefits
that renders it a versatile tool for a wide range of physical situations.
It can be set up for vertices defined both in the Matsubara or Keldysh
formalism \cite{Jakobs2007,Karrasch2010a,Sieberer2016}, allowing the treatment of systems in and out of equilibrium.
Unlike matrix-product state approaches \cite{Schollwoeck2011}, it is oblivious to the degree of
entanglement in the system. Noninteracting parts of the system (i.e.,
leads) can be integrated out exactly and simply yield additional self-energy
contributions for the interacting degrees of freedom. Since self-energies
are nothing but single-particle vertices, this strategy readily integrates
in the fRG formalism and allows for the treatment of open systems.

In this work, we devise various second-order fRG schemes for one-dimensional spinless fermions, document their capabilities and limitations, and discuss technical details of their implementation. Our results extend previous first order fRG-1 studies which addressed transport
through nanowires \cite{Meden2003,Enss2005} as well as Luttinger liquid power laws in the presence of impurities \cite{Meden2002,Andergassen2004}
or disorder \cite{Karrasch2015}. A real-space fRG-2 framework for spinful one-dimensional fermions has been established recently and featured prominently in comparison to experimental data on the 0.7-anomaly in quantum transport \cite{Bauer2013,Bauer2014,Weidinger2017}. First simple attempts to devise a k-space fRG-2 scheme for the spinful Hubbard model can be found in Refs.~\cite{Honerkamp2003} and \cite{Tam2006}; the methods were used to compute the quasiparticle weight \cite{Honerkamp2003} and the ground state phase diagram \cite{Tam2006}. In the realm of quantum impurity problems, both fRG-1 as well as full fRG-2 approaches already exist \cite{Karrasch2008}. Momentum space fRG was widely used to investigate the flow of the four-point vertex for two-dimensional problems \cite{Halboth2000,Honerkamp2001,Salmhofer2004,Gersch2005}; the complexity of these systems, however, renders it very difficult to implement true fRG-2 schemes. The key goal of this paper is to develop and benchmark second-order fRG approaches to spinless systems in real- and k-space.

Our Hamiltonian takes the form
\begin{equation}
H=\sum_{j}-t_{j}c_{j}^{\dagger}c_{j+1}-h.c.+U_{j}(n_{j}-\frac{1}{2})(n_{j+1}-\frac{1}{2}),\label{eq:H}
\end{equation}
where $c_{j}^{\dagger}$ creates a fermion at site $j$ (we put the lattice constant to unity, $a=1$), and $n_{j}=c_{j}^{\dagger}c_{j}$
is the number operator. We allow for a spatial dependence of the nearest neighbor hoppings $t_j$ and interactions $U_j$ in order to address both translationally-invariant and inhomogenous systems. We focus on half-filling and ground state
properties (i.e., equilibrium at $T=0$). The model, which is related
to the XXZ-spin-$1/2$ chain by a Jordan-Wigner transformation, is well studied. In the homogeneous case, it is exactly solvable
by the Bethe-Ansatz, and its low-energy physics is governed by the Luttinger liquid fixed point for $U/t<2$ \cite{Giamarchi2003}; it is generally susceptible to matrix-product state
type methods like density matrix renormalization group (DMRG). The model thus represents an ideal testbed for fRG
method development, which is our central goal in this paper.

In order to investigate both translationally-invariant and inhomogenous systems, we use a Matsubara frequency cutoff (we briefly comment on results obtained via a k-space cutoff for the former case). The bare Green function $G_{0}\left(\omega\right)=[i\omega-H_{0}]^{-1}$
with $H_{0}$ representing the single-particle terms is augmented
by a sharp multiplicative cutoff
\begin{equation}
G_{0}^{\Lambda}\left(\omega\right)=\Theta\left(|\omega|-\Lambda\right)G_{0}\left(\omega\right).\label{G0Lam}
\end{equation}

The paper is organized as follows: In Sec. \ref{sec:r-Space} we develop
fRG-2 in real space as applicable for the inhomogeneous model as in
(\ref{eq:H}). As an application, we consider the backscattering of
carriers at an abrupt junction between two different but homogeneous
chains. In Sec. \ref{sec:k-space} we consider the homogeneous case
which allows us to work in k-space. Here, we use our formalism to
calculate the ground state occupation numbers and confirm Luttinger liquid
power laws with exponents of second order in the interaction strength.
We conclude in Sec. \ref{sec:Conclusion}. 

\section{\label{sec:r-Space}FRG in real space}

\paragraph{fRG flow equations.}

For the spinless fermion chain (\ref{eq:H}), the single-particle
indices (site index, Matsubara frequency) can be summarized in a multi-index
as $1=(j_{1},\omega_{1})$. Then, the fRG flow equation for the flowing
self-energy $\Sigma^{\Lambda}$ reads \cite{Metzner2012}
\begin{equation}
\partial_{\Lambda}\Sigma^{\Lambda}\left(1^{\prime};1\right)=-\sum_{2,2^{\prime}}S_{2,2^{\prime}}^{\Lambda}U^{\Lambda}\left(2^{\prime}1^{\prime};21\right)\label{eq:Sigma_flow}
\end{equation}
where $S^{\Lambda}=G^{\Lambda}\left(\partial_{\Lambda}\left[G_{0}^{\Lambda}\right]^{-1}\right)G^{\Lambda}$
is the single-scale propagator, and $G^{\Lambda}=[\left[G_{0}^{\Lambda}\right]^{-1}-\Sigma^{\Lambda}]^{-1}$. In a truncation neglecting
the 3-particle vertex (which is generated along the flow by terms
of third order in the bare interaction $v$), the flow equations for the
2-particle vertex $U^{\Lambda}$ read\cite{Metzner2012}
\begin{eqnarray}
\partial_{\Lambda}U_{\Pi}^{\Lambda}\left(1^{\prime}2^{\prime};12\right) & = & \sum_{33^{\prime}44^{\prime}}U^{\Lambda}\left(1^{\prime}2^{\prime};34\right)U^{\Lambda}\left(3^{\prime}4^{\prime};12\right)\label{eq:UPi_flow}\\
 &  & \times S_{33^{\prime}}^{\Lambda}G_{44^{\prime}}^{\Lambda}\nonumber 
\end{eqnarray}
\begin{eqnarray}
\partial_{\Lambda}U_{\Xi}^{\Lambda}\left(1^{\prime}2^{\prime};12\right) & = & \sum_{33^{\prime}44^{\prime}}U^{\Lambda}\left(1^{\prime}4^{\prime};32\right)U^{\Lambda}\left(3^{\prime}2^{\prime};14\right)\label{eq:UXi_flow}\\
 &  & \times[S_{33^{\prime}}^{\Lambda}G_{44^{\prime}}^{\Lambda}+S_{44^{\prime}}^{\Lambda}G_{33^{\prime}}^{\Lambda}],\nonumber 
\end{eqnarray}
\begin{eqnarray}
\partial_{\Lambda}U_{\Delta}^{\Lambda}\left(1^{\prime}2^{\prime};12\right) & = & \!-\!\sum_{33^{\prime}44^{\prime}}U^{\Lambda}\left(2^{\prime}4^{\prime};32\right)U^{\Lambda}\left(3^{\prime}1^{\prime};14\right)\label{eq:UDe_flow}\\
 &  & \times[S_{33^{\prime}}^{\Lambda}G_{44^{\prime}}^{\Lambda}+S_{44^{\prime}}^{\Lambda}G_{33^{\prime}}^{\Lambda}],\nonumber 
\end{eqnarray}
where we have introduced the notation
\begin{equation}
U^{\Lambda}=v+U_{\Pi}^{\Lambda}+U_{\Xi}^{\Lambda}+U_{\Delta}^{\Lambda}.
\end{equation}
Here, $v\left(1^{\prime}2^{\prime};12\right)$ is the bare antisymmetric
vertex, defined from the interacting Hamiltonian as $H_{I}=\frac{1}{4}\sum_{1,2,1^{\prime},2^{\prime}}v\left(1^{\prime}2^{\prime};12\right)c_{1^{\prime}}^{\dagger}c_{2^{\prime}}^{\dagger}c_{2}c_{1}$.
The following entries take the value $U_{j}$, independent of the
frequencies (for all vertex functions, conservation of Matsubara frequency
is implied)
\begin{eqnarray}
 &  & v\left(j,j+1;j,j+1\right)\\
 & = & v\left(j+1,j;j+1,j\right)\nonumber \\
 & = & -v\left(j,j+1;j+1,j\right)\nonumber \\
 & = & -v\left(j+1,j;j,j+1\right)\nonumber \\
 & = & U_{j}\nonumber 
\end{eqnarray}
and all $v$-entries of a different form vanish.

\paragraph{fRG-2 scheme, general form.}

The only approximation so far was to set the three-particle vertex to zero. However, the number of remaining flow equations generally scales with the fourth power of the number of interacting degrees of freedom; a full solution is therefore only possible for impurity problems \cite{Karrasch2008}, and further simplifications need to be devised \cite{Bauer2014} for 1d systems.

A straightforward way to set up a fRG-2 scheme is to approximate $U^{\Lambda}\rightarrow v$
on the right hand side of Eqs.~(\ref{eq:UPi_flow}) to (\ref{eq:UDe_flow}).
This choice greatly simplifies the frequency dependence of $U_{\Pi,\Xi,\Delta}^{\Lambda}\left(1^{\prime}2^{\prime};12\right)$.
Each vertex depends on the correspondingly labeled composite (bosonic)
Matsubara frequency
\begin{eqnarray}
\Pi & = & \omega_{1^{\prime}}+\omega_{2^{\prime}}=\omega_{1}+\omega_{2},\label{eq:PI}\\
\Xi & = & \omega_{2^{\prime}}-\omega_{1}=\omega_{2}-\omega_{1^{\prime}},\label{eq:Xi}\\
\Delta & = & \omega_{1^{\prime}}-\omega_{1}=\omega_{2}-\omega_{2^{\prime}},\label{eq:Delta}
\end{eqnarray}
respectively. Note that we explicitly used fermionic frequency conservation, $\omega_{1}^\prime+\omega_{2}^\prime=\omega_{1}+\omega_{2}$. At first glance it seems that by inserting the bare vertex on the right hand side of
Eqs. (\ref{eq:UPi_flow}) to (\ref{eq:UDe_flow}), one generates a large number of different index structures on the left
hand side. However, it turns out that many of them are related by
symmetries (time-reversal, complex-conjugation or anti-symmetry) of
the vertex such as $U_{\Delta}^{\Lambda}\left(1^{\prime}2^{\prime};12\right)=-U_{\Xi}^{\Lambda}\left(2^{\prime}1^{\prime};12\right)$.

\paragraph{fRG-2 scheme, tight-binding chain.}

For the tight-binding chain defined in Eq. (\ref{eq:H}), the set of independent index structures is given by\begin{widetext}
\begin{eqnarray}
U_{\Pi}^{\Lambda}\left(j_{1},\Pi-\omega^{\prime};j_{1}+1,\omega^{\prime};j_{2},\Pi-\omega;j_{2}+1,\omega\right) & \equiv & P_{j_{1}j_{2}}^{++}\left(\Pi\right),\\
U_{\Xi}^{\Lambda}\left(j_{2},\omega^{\prime};j_{1},\Xi+\omega;j_{1},\omega;j_{2},\Xi+\omega^{\prime}\right) & \equiv & X_{j_{1}j_{2}}^{00}\left(\Xi\right),\\
U_{\Xi}^{\Lambda}\left(j_{2},\omega^{\prime};j_{1}-1,\Xi+\omega;j_{1},\omega;j_{2},\Xi+\omega^{\prime}\right) & \equiv & X_{j_{1}j_{2}}^{0-}\left(\Xi\right),\\
U_{\Xi}^{\Lambda}\left(j_{2}+1,\omega^{\prime};j_{1}+1,\Xi+\omega;j_{1},\omega;j_{2},\Xi+\omega^{\prime}\right) & \equiv & X_{j_{1}j_{2}}^{++}\left(\Xi\right),\\
U_{\Xi}^{\Lambda}\left(j_{2}+1,\omega^{\prime};j_{1}-1,\Xi+\omega;j_{1},\omega;j_{2},\Xi+\omega^{\prime}\right) & \equiv & X_{j_{1}j_{2}}^{+-}\left(\Xi\right),
\end{eqnarray}
which can be conveniently treated as matrices depending on a single
frequency. Their flow equations read
\begin{eqnarray}
\partial_{\Lambda}P_{j_{1}j_{2}}^{++}\left(\Pi\right) & = & T\sum_{\omega}U_{j_{1}}[S_{j_{1}j_{2}}^{\Lambda}\left(\omega\right)G_{j_{1}+1,j_{2}+1}^{\Lambda}\left(\Pi-\omega\right)+S_{j_{1}+1j_{2}+1}^{\Lambda}\left(\omega\right)G_{j_{1},j_{2}}^{\Lambda}\left(\Pi-\omega\right)\label{eq:P++}\\
 &  & -S_{j_{1}+1j_{2}}^{\Lambda}\left(\omega\right)G_{j_{1},j_{2}+1}^{\Lambda}\left(\Pi-\omega\right)-S_{j_{1}j_{2}+1}^{\Lambda}\left(\omega\right)G_{j_{1}+1,j_{2}}^{\Lambda}\left(\Pi-\omega\right)]U_{j_{2}},\nonumber \\
\partial_{\Lambda}X_{j_{1}j_{2}}^{00}\left(\Xi\right) & = & T\sum_{\omega}[U_{j_{2}}U_{j_{1}}S_{j_{2}+1,j_{1}+1}^{\Lambda}\left(\omega\right)G_{j_{1}+1,j_{2}+1}^{\Lambda}\left(\omega-X\right)+U_{j_{1}-1}U_{j_{2}}S_{j_{2}+1,j_{1}-1}^{\Lambda}\left(\omega\right)G_{j_{1}-1,j_{2}+1}^{\Lambda}\left(\omega-X\right)\label{eq:X00}\\
 &  & +U_{j_{1}}U_{j_{2}-1}S_{j_{2}-1j_{1}+1}^{\Lambda}\left(\omega\right)G_{j_{1}+1,j_{2}-1}^{\Lambda}\left(\omega-X\right)+U_{j_{1}-1}U_{j_{2}-1}S_{j_{2}-1j_{1}-1}^{\Lambda}\left(\omega\right)G_{j_{1}-1,j_{2}-1}^{\Lambda}\left(\omega-X\right)]+S\leftrightarrow G,\nonumber 
\end{eqnarray}
\begin{eqnarray}
\partial_{\Lambda}X_{j_{1}j_{2}}^{0-}\left(\Xi\right) & = & -T\sum_{\omega}[U_{j_{1}-1}U_{j_{2}}G_{j_{1}-1,j_{2}+1}^{\Lambda}\left(\omega-X\right)S_{j_{2}+1,j_{1}}^{\Lambda}\left(\omega\right)+U_{j_{1}-1}U_{j_{2}-1}G_{j_{1}-1,j_{2}-1}^{\Lambda}\left(\omega-X\right)S_{j_{2}-1j_{1}}^{\Lambda}\left(\omega\right)\label{eq:X0-}\\
 &  & +U_{j_{1}-1}U_{j_{2}}S_{j_{1}-1,j_{2}+1}^{\Lambda}\left(\omega\right)G_{j_{2}+1,j_{1}}^{\Lambda}\left(\omega+X\right)+U_{j_{1}-1}U_{j_{2}-1}S_{j_{1}-1,j_{2}-1}^{\Lambda}\left(\omega\right)G_{j_{2}-1j_{1}}^{\Lambda}\left(\omega+X\right)],\nonumber \\
\partial_{\Lambda}X_{j_{1}j_{2}}^{++}\left(\Xi\right) & = & T\sum_{\omega}U_{j_{1}}\left(S_{j_{2}+1j_{1}}^{\Lambda}\left(\omega\right)G_{j_{1}+1,j_{2}}^{\Lambda}\left(\omega-X\right)+S_{j_{1}+1,j_{2}}^{\Lambda}\left(\omega\right)G_{j_{2}+1j_{1}}^{\Lambda}\left(\omega+X\right)\right)U_{j_{2}},\label{eq:X++}\\
\partial_{\Lambda}X_{j_{1}j_{2}}^{+-}\left(\Xi\right) & = & T\sum_{\omega}U_{j_{1}-1}\left(S_{j_{2}+1j_{1}}^{\Lambda}\left(\omega\right)G_{j_{1}-1,j_{2}}^{\Lambda}\left(\omega-X\right)+S_{j_{1}-1,j_{2}}^{\Lambda}\left(\omega\right)G_{j_{2}+1j_{1}}^{\Lambda}\left(\omega+X\right)\right)U_{j_{2}}.\label{eq:X+-}
\end{eqnarray}
Expressed in terms of these quantities, the flow equation for the self-energy
(\ref{eq:Sigma_flow}) takes the form
\begin{eqnarray}
\partial_{\Lambda}\Sigma_{ij}^{\Lambda}\left(\nu\right) & = & T\sum_{\omega}\delta_{ij}\left[-U_{i-1}S_{i-1,i-1}^{\Lambda}\left(\omega\right)-U_{i}S_{i+1,i+1}^{\Lambda}\left(\omega\right)\right]+\delta_{i+1j}U_{i}S_{i,j}^{\Lambda}\left(\omega\right)+\delta_{i-1,j}S_{i,j}^{\Lambda}\left(\omega\right)U_{j}\label{eq:Sflow_final}\\
 &  & +\delta_{ij}\sum_{k}\left(S_{kk}^{\Lambda}\left(\omega\right)X_{kj}^{00}\left(0\right)+S_{k,k-1}^{\Lambda}\left(\omega\right)X_{kj}^{0-}\left(0\right)+S_{k,k+1}^{\Lambda}\left(\omega\right)X_{k+1,j}^{0-}\left(0\right)\right)\nonumber \\
 &  & +\delta_{i,j-1}\sum_{k}\left(S_{kk}^{\Lambda}\left(\omega\right)X_{jk}^{0-}\left(0\right)+S_{k,k-1}^{\Lambda}\left(\omega\right)X_{k-1,j-1}^{++}\left(0\right)+S_{k,k+1}^{\Lambda}\left(\omega\right)X_{jk}^{+-}\left(0\right)\right)\nonumber \\
 &  & +\delta_{i,j+1}\sum_{k}\left(S_{kk}^{\Lambda}\left(\omega\right)X_{j+1,k}^{0-}\left(0\right)+S_{k,k-1}^{\Lambda}\left(\omega\right)X_{kj}^{+-}\left(0\right)+S_{k,k+1}^{\Lambda}\left(\omega\right)X_{kj}^{++}\left(0\right)\right)\nonumber \\
 &  & -S_{j-1,i-1}^{\Lambda}\left(\omega\right)P_{i-1,j-1}^{++}\left(\nu+\omega\right)-S_{i+1,j-1}^{\Lambda}\left(\omega\right)P_{j-1,i}^{++}\left(\nu+\omega\right)-S_{i-1,j+1}^{\Lambda}\left(\omega\right)P_{i-1,j}^{++}\left(\nu+\omega\right)\nonumber \\
 &  & +S_{i+1,j+1}^{\Lambda}\left(\omega\right)P_{i,j}^{++}\left(\nu+\omega\right)+S_{i,j+1}^{\Lambda}\left(\omega\right)X_{j+1,i}^{0-}\left(\nu-\omega\right)+S_{i-1,j-1}^{\Lambda}\left(\omega\right)X_{j,i-1}^{+-*}\left(\nu-\omega\right)\nonumber \\
 &  & -S_{ij}^{\Lambda}\left(\omega\right)X_{ij}^{00}\left(\nu-\omega\right)+S_{i+1,j}^{\Lambda}\left(\omega\right)X_{i+1j}^{0-}\left(\nu-\omega\right)+S_{i,j-1}^{\Lambda}\left(\omega\right)X_{ji}^{0-*}\left(\nu-\omega\right)+S_{i-1,j}^{\Lambda}\left(\omega\right)X_{i,j}^{0-*}\left(\nu-\omega\right)\nonumber \\
 &  & -S_{i-1,j+1}^{\Lambda}\left(\omega\right)X_{i-1j}^{++}\left(\nu-\omega\right)+S_{i+1,j-1}^{\Lambda}\left(\omega\right)X_{i,j-1}^{++*}\left(\nu-\omega\right)+S_{i+1,j+1}^{\Lambda}\left(\omega\right)X_{i+1,j}^{+-}\left(\nu-\omega\right).\nonumber 
\end{eqnarray}

\end{widetext} Concerning initial conditions, the single-particle
contributions $U_{j}n_{j}$ and $U_{j+1}n_{j}$ in (\ref{eq:H}) are
the boundary condition for $\Sigma^{\infty}$. However, the flow from
$\Lambda=\infty$ to finite but large $\Lambda_{i}$ can be integrated
analytically \cite{Karrasch2008}. Effectively, by using Eq. (\ref{eq:Sigma_flow}),
the initial conditions at the start of the numerically integrated
flow at $\Lambda_{i}$ read $U_{\Pi,\Xi,\Delta}^{\Lambda_{i}}=0$
and $\Sigma^{\Lambda_{i}}=0$.

In the zero temperature case that we are interested in exclusively,
$\omega$ is continuous. We find for the single-scale propagator and
its product with $G^{\Lambda}$ \cite{Morris1993,Karrasch2008}
\begin{align}
S^{\Lambda}(\omega) & =\delta\!\left(|\omega|-\Lambda\right)\tilde{G}^{\Lambda}(\omega)\label{eq:S(T=00003D0)}\\
S^{\Lambda}\left(\omega\right)G^{\Lambda}\left(\nu\right) & =\delta\!\left(|\omega|-\Lambda\right)\Theta\!\left(|\nu|-\Lambda\right)\tilde{G}^{\Lambda}\left(\omega\right)\tilde{G}^{\Lambda}\left(\nu\right)\label{eq:SG(T=00003D0)}
\end{align}
where $\Theta(x)$ is a unit-step function with $\Theta(0)=1/2$ and
$\tilde{G}^{\Lambda}(\omega)=\left[G_{0}^{-1}(\omega)-\Sigma^{\Lambda}(\omega)\right]^{-1}$.
The $\delta$-function conveniently cancels the frequency integral
$T\sum_{\omega}\rightarrow\frac{1}{2\pi}\int_{-\infty}^{\infty}d\omega$
so that we are left with sums of the form $\sum_{\omega=\pm\Lambda}$.

To summarize, our fRG-2 scheme for the spinless fermion model (\ref{eq:H})
consists of a solution of the system of differential equations (\ref{eq:P++})
to (\ref{eq:Sflow_final}) with the replacements (\ref{eq:S(T=00003D0)})
and (\ref{eq:SG(T=00003D0)}) appropriate for $T=0$ and the boundary
conditions discussed above. We remark that our fRG-2 scheme can be derived as a special case from the more general scheme in Ref. \cite{Weidinger2017} which was set up for range-$L$ correlations and spinful models. In particular, one would have to project the flow eqs. (29) and (33) in Ref. \cite{Weidinger2017} to a single spin component, let $L=1$ and neglect any feedback between vertices.

\paragraph{Details of numerical implementation.}

On the technical level, our numerical code makes extensive use of
fast matrix-matrix and matrix-vector multiplications to evaluate the
right hand side of the flow equations. We used a Runge-Kutta integration
routine and discretize the continuous Matsubara frequencies on a geometric
grid with $\mathcal{O}(100)$ points ranging from $\omega_{min}=\mathcal{O}(10^{-6}t)$
to $\omega_{max}=\mathcal{O}(1000t)\simeq\Lambda_{i}$ (in units of
typical hopping $t$). We avoid negative Matsubara frequencies by
using time-reversal symmetry for vertices and self-energies. Linear
interpolation between grid-point data is used on the right-hand side
of the flow equations. With these choices, we are able to integrate
the flow equations for systems consisting of $\mathcal{O}(100)$ interacting
sites on a standard desktop computer within a few hours.

Due to the complexity of the flow equations, we have performed tests
against exact diagonalization (ED) for small systems (with
constant $U_{j}=U$). We compared single particle ground state expectation
values $\left\langle c_{i}^{\dagger}c_{j}\right\rangle $ and confirmed
$\left\langle c_{i}^{\dagger}c_{j}\right\rangle _{FRG}-\left\langle c_{i}^{\dagger}c_{j}\right\rangle _{ED}=\mathcal{O}\left(U^{3}\right)$
consistent with the claim that our fRG scheme is correct up to second
order. Here, 
\begin{align}
\left\langle c_{i}^{\dagger}c_{j}\right\rangle _{FRG} & =\frac{1}{2\pi}\int_{-\infty}^{\infty}d\omega\,e^{i\omega\eta}\,G_{ij}^{\Lambda=0}\left(\omega\right)\label{eq:occ}\\
 & =\frac{1}{2}\delta_{ij}+\frac{1}{2\pi}\int_{-\Lambda_{i}}^{\Lambda_{i}}d\omega\,G_{ij}^{\Lambda=0}\left(\omega\right),\nonumber 
\end{align}
where in the second step the large-$\omega$ tails of the Greens function
have been treated analytically to produce the leading term in the
last line. 

It is possible to go beyond the second order scheme presented above
without increasing the complexity of the presented vertex structures.
For this, one can feed back parts of $U_{\Pi,\Xi,\Delta}^{\Lambda}(0)$
on the right-hand side of Eqs. (\ref{eq:UPi_flow}) to (\ref{eq:UDe_flow})
that resemble the spatial index structure of $v$, see Ref. \cite{Bauer2014}.
However, implementing this partial feedback procedure favoring only
nearest-neighbor interactions, we observed no systematic improvement
in comparison to various reference data from exact diagonalization.
Consequently, we discard this feedback option in the following.

\paragraph{Application: Conductance across abrupt junctions.}

We now turn to an application demonstrating the performance of the
above fRG-2 scheme for spinless fermions. Whereas corresponding first
order schemes (or variants thereof including a subset of higher order
terms) have been employed to demonstrate Luttinger liquid physics
by reproducing analytically known power laws which manifest for very large systems
of $\mathcal{O}(10^{5})$ sites \cite{Andergassen2004}, the objectives
for the second order scheme must be different due to the inherent
limitation on system sizes (we will investigate power laws in momentum space in the next section).
\begin{figure}
\noindent \begin{centering}
\includegraphics{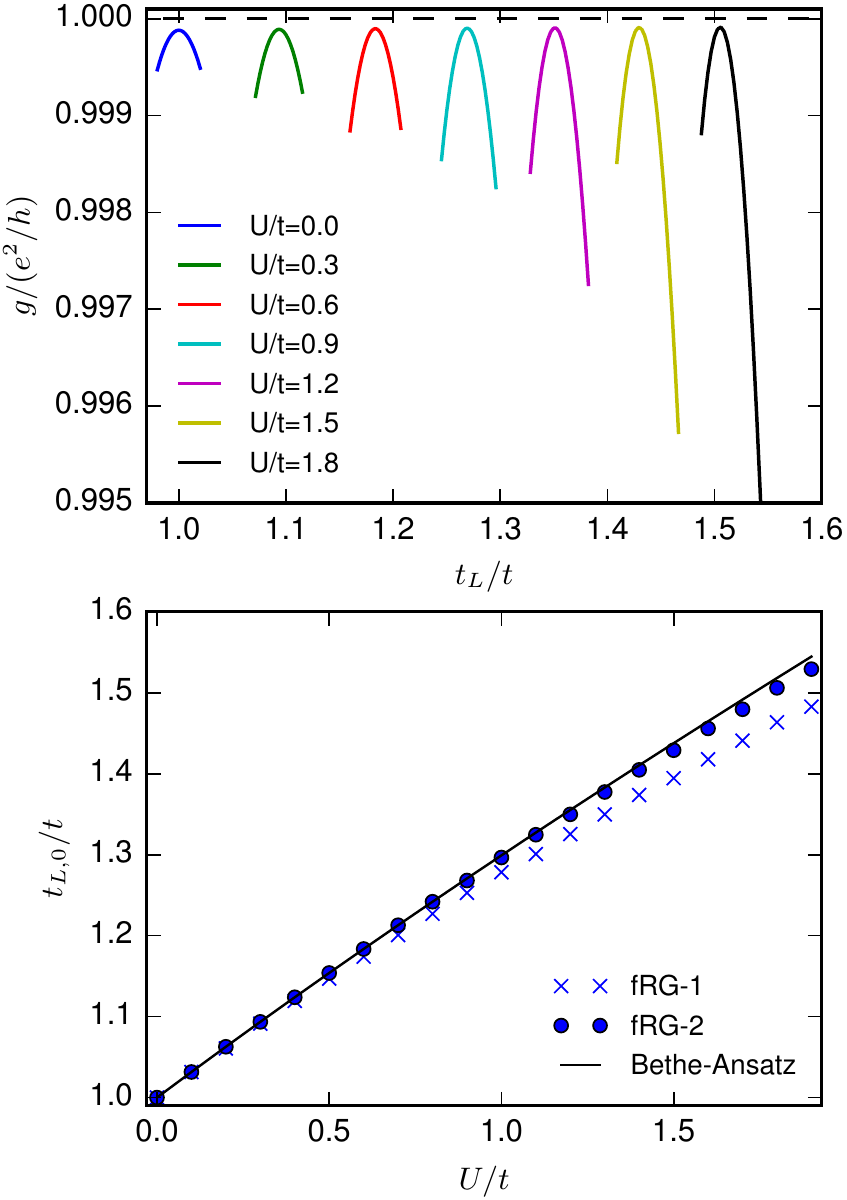}
\par\end{centering}
\caption{\label{fig:PerfectG}Top panel: Conductance $g$ across an interacting
region of length $L=120$ of a spinless fermion model (Eq. (\ref{eq:H})
with $t_{j}=t$ and $U_{j}=U$) connected to non-interacting leads
characterized by hoppings $t_{L}$ as simulated with fRG-2. Various
$U$, increasing from left to right, are depicted. Bottom panel: The
conductance maxima for different $U$ occur at values $t_{L,0}^{fRG-2}$
(dots) that nicely agree with the analytical predictions based
on Luttinger liquid theory from Eq. (\ref{eq:tL for perfect G}) (solid
line). Analogous results from fRG-1 are shown as crosses. }
\end{figure}

We apply the fRG-2 scheme to a transport setup where the $t_{j}$
and $U_{j}$ in Eq. (\ref{eq:H}) are adjusted to form an interacting
region of length $L$ {[}modeled with $U_{j} \! = \! U \! > \! 0$ for $j=0,...,L-2${]}
which is abruptly connected to non-interacting leads on the left and
right site (with $U_{j} \! = \! 0$ for $j\! < \! 0$ or $j \! > \! L-2$). The hopping strength
in the interacting region is set to $t_{j}=t$. In the leads and their
connections to the interacting region, it is taken to be $t_{L}$. Generically,
backscattering occurs in such a situation at the abrupt contacts and completely suppresses transport through a Luttinger liquid in the zero-temperature, infinite-system limit. One would expect that transport is only possible if the abrupt junction is turned into a smooth adiabatic one  (with a then perfect conductance $g=e^{2}/h$) by 
changing the values of $t_{j}$ and $U_{j}$ gradually with $j$ \cite{Karrasch2015}. However, it was
analytically predicted by Sedlmayr et al. in Refs. \cite{Sedlmayr2012,Sedlmayr2014}
that backscattering can also be avoided in abrupt junctions if
the velocity of the elementary (bulk-)excitations match across the
two connected regions ({see also \cite{Janzen2006} for related ideas). For a homogeneous system as in Eq. (\ref{eq:H})
with $t_{j}=t$ and $U_{j}=U$, the velocity (for $|U|<2t$) is known
to be \cite{Giamarchi2003},
\begin{equation}
u=t\pi\frac{\sqrt{1-U^{2}/4t^{2}}}{\arccos\left(U/2t\right)}\label{eq:u}
\end{equation}
from the Bethe-Ansatz solution. Accordingly, we expect perfect conductance
for given $t,U$ once $t_{L}$ is tuned to 
\begin{equation}
t_{L,0}^{Bethe}=t\sqrt{1-U^{2}/4t^{2}}/\arccos\left(U/2t\right).\label{eq:tL for perfect G}
\end{equation}
This analytical prediction has only recently been tested numerically
in Ref. \cite{Morath2016} on the basis of a numerical conductance
computation: Using a Monte-Carlo simulation the conductance $g$ across
a single junction was calculated as a function of $t_{L}$. However,
since the numerical method was plagued with large error bars on the
resulting conductances, a clear determination of the $t_{L}$ at which
a maximum of $g$ resulted was impossible (see Fig. 7 in Ref. \cite{Morath2016}),
and only a consistency check was achieved. 

In the following we show that the fRG-2 scheme is well-suited to
check the above prediction and give solid numerical evidence for its
validity. We implement the transport setup described above for $L=120$
and calculate the conductance via $g=4\frac{e^{2}}{h}t_{L}^{2}\left|G_{0,L-1}^{R}\left(0\right)\right|^{2}$
\cite{Karrasch2010} from the retarded Green function $G^{R}$. Note that at $T=0$, this expression for $g$ is exact since vertex corrections are absent \cite{Oguri2001}. In the fRG formalism,
the semi-infinite leads can be taken into account analytically by adding local self-energy terms $\Sigma_{j,j}\left(i\omega\right)=\frac{1}{2}\left(i\omega-i\sqrt{4t_{L}^{2}+\omega^{2}}\right)$ for $\omega>0$ to the sites $j=0$ and $j=L-1$
\cite{Karrasch2010}. 

The conductance $g(t_{L})$ as a function of the hopping $t_L$ in the leads is shown in Fig. \ref{fig:PerfectG} (top). One can clearly identify
maxima at $t_{L}\equiv t_{L,0}^{fRG-2}$; these values are extracted
and plotted vs. $U$ in the bottom panel (dots). Our results closely
match the analytical prediction $t_{L,0}^{Bethe}$ defined via Eq.
(\ref{eq:tL for perfect G}) (solid line) except at the largest $U$.
We have checked that while the value of the conductance maximum increases
towards unity with increasing $L$, its position $t_{L,0}^{fRG-2}$
does not change. Moreover, the crosses in Fig. \ref{fig:PerfectG}
(bottom) denote the corresponding results of a first order fRG-1 calculation
(using Eq. (\ref{eq:Sigma_flow}) with the bare interaction only)
showing larger deviations from the Bethe-Ansatz value for large
$U$.

\section{\label{sec:k-space}FRG in k-space}

\paragraph{fRG flow equations.}

We now devise a second-order fRG scheme in momentum space. Our testbed will be the translational invariant spinless fermion
chain at half-filling, defined by letting $t_{j}=t$ and $U_{j}=U$ in Hamiltonian
(\ref{eq:H}). After a Fourier transformation in the spatial index
of the fermion operators $c_{j}=\int_{k}c_{k}e^{ikj}$ (with $\int_{k}\equiv\frac{1}{2\pi}\int_{-\pi}^{\pi}dk$),
we obtain for the non-interacting part $H_{0}=\int_{k}c_{k}^{\dagger}H_{0}(k)c_{k}$
with
\begin{equation}
H_{0}\left(k\right)=-2t\cos\left(k\right),
\end{equation}
and the bare interaction vertex reads \cite{Shankar1994}
\begin{align}
 & v\left(k_{1}^{\prime},k_{2}^{\prime};k_{1},k_{2}\right)=\\
 & \bar{\delta}_{k_{1}^{\prime}+k_{2}^{\prime}-k_{1}-k_{2}}2U\left(\cos\left[k_{1}^{\prime}-k_{1}\right]-\cos\left[k_{2}^{\prime}-k_{1}\right]\right),\nonumber 
\end{align}
where $\bar{\delta}_{k_{1}^{\prime}+k_{2}^{\prime}-k_{1}-k_{2}}$
implies conservation of crystal momentum up to integer multiples of
$2\pi$. The flow equations simplify compared to Sec. \ref{sec:r-Space}
due to the Green functions being diagonal in momentum. For the self-energy
$\Sigma^{\Lambda}$ one finds
\begin{equation}\label{eq:flowsek}
\partial_{\Lambda}\Sigma^{\Lambda}\left(\begin{array}{c}
k_{1}\\
\omega_{1}
\end{array}\right)=-\int_{q}\int_{\omega}S^{\Lambda}\left(\begin{array}{c}
q\\
\omega
\end{array}\right)U^{\Lambda}\left(\begin{array}{ccc}
q & k_{1}; & q\\
\omega & \omega_{1}; & \omega
\end{array}\right),
\end{equation}
 and the 2-particle vertex $U^{\Lambda}$, in a truncation neglecting
the 3-particle vertex, flows according to\begin{widetext}
\begin{eqnarray}
 &  & \partial_{\Lambda}U^{\Lambda}\left(\begin{array}{ccc}
k_{1}^{\prime} & k_{2}^{\prime}; & k_{1}\\
\omega_{1}^{\prime} & \omega_{2}^{\prime}; & \omega_{1}
\end{array}\right)=\label{eq:DU_kSpace}\\
 & + & \int_{q}\int_{\omega}G^{\Lambda}\left(\begin{array}{c}
q\\
\omega
\end{array}\right)S^{\Lambda}\left(\begin{array}{c}
\left|k^{\prime}\right|\\
\omega^{\prime}
\end{array}\right)U^{\Lambda}\left(\begin{array}{ccc}
q & \left|k^{\prime}\right|; & k_{1}\\
\omega & \omega^{\prime}; & \omega_{1}
\end{array}\right)U^{\Lambda}\left(\begin{array}{ccc}
k_{1}^{\prime} & k_{2}^{\prime}; & q\\
\omega_{1}^{\prime} & \omega_{2}^{\prime}; & \omega
\end{array}\right)\nonumber \\
 & + & \int_{q}\int_{\omega}[G^{\Lambda}\left(\begin{array}{c}
q\\
\omega
\end{array}\right)S^{\Lambda}\left(\begin{array}{c}
\left|k^{\prime\prime}\right|\\
\omega^{\prime\prime}
\end{array}\right)+S^{\Lambda}\left(\begin{array}{c}
q\\
\omega
\end{array}\right)G^{\Lambda}\left(\begin{array}{c}
\left|k^{\prime\prime}\right|\\
\omega^{\prime\prime}
\end{array}\right)]U^{\Lambda}\left(\begin{array}{ccc}
k_{1}^{\prime} & \left|k^{\prime\prime}\right|; & q\\
\omega_{1}^{\prime} & \omega^{\prime\prime}; & \omega
\end{array}\right)U^{\Lambda}\left(\begin{array}{ccc}
q & k_{2}^{\prime}; & k_{1}\\
\omega & \omega_{2}^{\prime}; & \omega_{1}
\end{array}\right)\nonumber \\
 & - & \left\{ \omega_{1}^{\prime}\leftrightarrow\omega_{2}^{\prime},\,k_{1}^{\prime}\leftrightarrow k_{2}^{\prime}\right\} \nonumber 
\end{eqnarray}
where $k^{\prime}=k_{1}^{\prime}+k_{2}^{\prime}-q$, $k^{\prime\prime}=q+k_{2}^{\prime}-k_{1}$
and analogously for Matsubara frequencies $\omega^{\prime}=\omega_{1}^{\prime}+\omega_{2}^{\prime}-\omega$, $\, \omega^{\prime\prime}=\omega+\omega_{2}^{\prime}-\omega_{1}$.
Further, $|k|$ denotes the projection of $k$ to the first Brillouin
zone (BZ) $[-\pi,+\pi)$. We reduced the arguments of the vertex functions
to three momenta and frequencies by virtue of the momentum and frequency
conserving delta-functions. For the momenta, note that if we fix $k_{1}^{\prime},k_{2}^{\prime},k_{1}\in[-\pi,+\pi)$
in the first BZ, there is no ambiguity in the choice of $k_{2}\in[-\pi,+\pi)$
fulfilling crystal momentum conservation. We again switch from fermionic
to bosonic Matsubara frequencies $\Pi,\Xi,\Delta$ as above in Eqs.
(\ref{eq:PI}) to (\ref{eq:Delta}).

\paragraph{fRG-2 scheme, general form.}

As before, the general flow equations (\ref{eq:flowsek}) and (\ref{eq:DU_kSpace}) cannot be solved in practice, and we need to devise further approximations. The second-order approach which is equivalent to the one used in the real space
scheme amounts to replacing $U^{\Lambda}\rightarrow v$ on the
right hand side of Eq. (\ref{eq:DU_kSpace}). This yields a flowing
vertex of the form
\begin{eqnarray}
U^{\Lambda}\left(\begin{array}{ccc}
k_{1}^{\prime} & k_{2}^{\prime}; & k_{1}\\
\Pi & \Xi; & \Delta
\end{array}\right) & = & v\left(k_{1}^{\prime},k_{2}^{\prime};k_{1}\right)+U_{\Pi}^{\Lambda}\left(\begin{array}{cccc}
\Pi; & k_{1}^{\prime} & k_{2}^{\prime}; & k_{1}\end{array}\right)+U_{\Xi}^{\Lambda}\left(\begin{array}{cccc}
\Xi; & k_{1}^{\prime} & k_{2}^{\prime}; & k_{1}\end{array}\right)+U_{\Delta}^{\Lambda}\left(\begin{array}{cccc}
\Delta; & k_{1}^{\prime} & k_{2}^{\prime}; & k_{1}\end{array}\right).\label{eq:sum-ansatz}
\end{eqnarray}
As in the real space scheme in Sec. \ref{sec:r-Space}, the flow of
the three terms $U_{\Pi}^{\Lambda}$, $U_{\Xi}^{\Lambda}$ and $U_{\Delta}^{\Lambda}$
can be defined by the first, second and third line of the vertex flow
equation with $U^{\Lambda}\rightarrow v$ on the right hand side,
respectively.

Instead of stopping here (as we did in the real space scheme),
we want to take advantage of the simplifying assumption of translational
invariance to devise a meaningful feedback scheme. Note that a full
feedback of $U^{\Lambda}$ would yield a vertex that depends on frequencies
in a generic fashion and thus spoils the simple frequency structure
of Eq. (\ref{eq:sum-ansatz}). Although the parametrization of the
vertex $U^{\Lambda}$ using three frequency variables was implemented
for zero-dimensional models like the single-impurity Anderson model
in Ref. \cite{Karrasch2008}, the need for a dense grid of k points
in the vicinity of the Fermi points $k_F=\pm\pi/2$ requires a reduction
of the vertex complexity in the frequency variables to ensure applicability.
Here, while sticking to the form (\ref{eq:sum-ansatz})
as an \emph{Ansatz} for $U^{\Lambda}$ \cite{Karrasch2008,Bauer2014},
we observe that it consistently allows for a feedback of its static
part $U_{(0)}^{\Lambda}$, specifically
\begin{equation}
U_{(0)}^{\Lambda}=U_{0}+U_{\Pi}^{\Lambda}\left(\Pi=0\right)+U_{\Xi}^{\Lambda}\left(\Xi=0\right)+U_{\Delta}^{\Lambda}\left(\Delta=0\right).\label{eq:U(0)}
\end{equation}
Thus, a full feedback is established in terms of the vertex momentum dependence and is completely absent for the frequency dependence. This choice is
partially motivated by analogy to the structure of the bare vertex,
which has a nontrivial momentum- but trivial frequency dependence.
On the other hand, it turns out that a feedback of the spatial features
of the vertex is both necessary and sufficient to capture the phase
transition to the charge-density wave long-range ordered phase for
$U>2t$ \cite{Markhof}. With the feedback (\ref{eq:U(0)}), the flow equations
in the $T=0$ Matsubara cutoff scheme read
\begin{eqnarray}
\partial_{\Lambda}\Sigma^{\Lambda}\left(\begin{array}{c}
k_{1}\\
\omega_{1}
\end{array}\right) & = & -\frac{\sum_{\Omega=\pm\Lambda}}{\left(2\pi\right)}\int_{q}\tilde{G}^{\Lambda}\left(\begin{array}{c}
k\\
\Omega
\end{array}\right)\left[U_{0}+U_{\Pi}^{\Lambda}\left(\omega_{1}+\Omega\right)+U_{\Xi}^{\Lambda}\left(\omega_{1}-\Omega\right)+U_{\Delta}^{\Lambda}\left(0\right)\right]\left(\begin{array}{ccc}
q & k_{1}; & q\end{array}\right)\label{eq:flow_kSpace_Sigma}\\
\partial_{\Lambda}U_{\Pi}^{\Lambda}\left(\!\begin{array}{cccc}
\Pi; & \!k_{1}^{\prime} & \!k_{2}^{\prime}; & \!k_{1}\!\end{array}\right) & = & \frac{\sum_{\Omega=\pm\Lambda}}{\left(2\pi\right)}\int_{q}\theta_{\Lambda}\left(\Omega+\Pi\right)\tilde{G}^{\Lambda}\left(\begin{array}{c}
q\\
\Omega+\Pi
\end{array}\right)\tilde{G}^{\Lambda}\left(\begin{array}{c}
\left|k^{\prime}\right|\\
-\Omega
\end{array}\right)U_{(0)}^{\Lambda}\left(\begin{array}{ccc}
k_{1}^{\prime} & k_{2}^{\prime}; & q\end{array}\right)U_{(0)}^{\Lambda}\left(\begin{array}{ccc}
q & \left|k^{\prime}\right|; & k_{1}\end{array}\right)\label{eq:flow_kSpace_UPi}\\
\partial_{\Lambda}U_{\Xi}^{\Lambda}\left(\!\begin{array}{cccc}
\Xi; & \!k_{1}^{\prime} & \!k_{2}^{\prime}; & \!k_{1}\!\end{array}\right) & = & \frac{\sum_{\Omega=\pm\Lambda}}{\left(2\pi\right)}\int_{q}\left[\theta_{\Lambda}\left(\Omega-\Xi\right)\tilde{G}^{\Lambda}\left(\begin{array}{c}
q\\
\!\Omega-\Xi\!
\end{array}\right)\!\tilde{G}^{\Lambda}\left(\begin{array}{c}
\!\left|k^{\prime\prime}\right|\!\\
\Omega
\end{array}\right)\!+\!\theta_{\Lambda}\left(\Omega+\Xi\right)\tilde{G}^{\Lambda}\left(\begin{array}{c}
\!q\!\\
\!\Omega\!
\end{array}\right)\!\tilde{G}^{\Lambda}\left(\begin{array}{c}
\!\left|k^{\prime\prime}\right|\!\\
\!\Omega+\Xi\!
\end{array}\right)\right]\label{eq:flow_kSpace_UXi}\\
 &  & \times U_{(0)}^{\Lambda}\left(\begin{array}{ccc}
q & k_{2}^{\prime}; & k_{1}\end{array}\right)U_{(0)}^{\Lambda}\left(\begin{array}{ccc}
k_{1}^{\prime} & \left|k^{\prime\prime}\right|; & q\end{array}\right)\nonumber 
\end{eqnarray}
where $\theta_{\Lambda}\left(\Omega\right)\equiv\Theta\left(|\Omega|-\Lambda\right)$
and $\partial_{\Lambda}U_{\Xi}^{\Lambda}\left(\begin{array}{cccc}
\nu; & k_{1}^{\prime} & k_{2}^{\prime}; & k_{1}\end{array}\right)=-\partial_{\Lambda}U_{\Delta}^{\Lambda}\left(\begin{array}{cccc}
\nu; & k_{2}^{\prime} & k_{1}^{\prime}; & k_{1}\end{array}\right)$. As in Sec. \ref{sec:r-Space}, the initial conditions are $\Sigma^{\Lambda_{i}}=0$
and $U_{\Pi,\Xi,\Delta}^{\Lambda_{i}}=0$. We emphasize that in terms
of perturbation theory, the solution of these flow equations is
correct to second order in the bare interaction $U$; the
k-space fRG proposed belongs to the class fRG-2, regardless of the
chosen feedback scheme. Also note that the equations presented in this section describe the flow of any translationally-invariant 1d system. The specific choice of (\ref{eq:H}) only enters during the evaluation of $\tilde{G}^{\Lambda}$ as well as in the initial condition. \end{widetext}

\paragraph{Details of numerical implementation.}

Before we turn to the presentation of results, let us describe some
main technical aspects of the numerical implementation of our fRG
scheme. As in Sec. \ref{sec:r-Space}, we use time-reversal symmetry
to relate vertices at negative Matsubara frequencies $-\omega$ to
positive $\omega$, $U_{\Omega}^{\Lambda}\left(\begin{array}{cccc}
\omega; & k_{1}^{\prime} & k_{2}^{\prime}; & k_{1}\end{array}\right)=U_{\Omega}^{\Lambda}\left(\begin{array}{cccc}
-\omega; & -k_{1}^{\prime} & -k_{2}^{\prime}; & -k_{1}\end{array}\right)^{\star}$ for $\Omega\in\{\Pi,X,D\}$ and the Matsubara frequencies are discretized
on a grid as in Sec. \ref{sec:r-Space}. Similarly, the momenta are
discretized with logarithmic precision around the Fermi points with
distances $\delta k_{n}=a^{n}\delta k_{min}$, with minimal distance
$\delta k_{min}=0.02$ and $a=1.5$. Towards $k=\pm\pi$ and $k=0$
the grid blends towards linear spacing. Finally, for the remaining
$q$-integrals on the right hand side of the flow equation, we work
with trapezoidal numerical integration on a $q$-grid similar (but
much finer) than the $k$-grid with highest resolution at the points
where the Green functions peak. 

On the right hand side of the flow equations, vertices and self-energies
need to be computed on generic values of frequencies and momenta that
do not coincide with values on the grid. In these cases, linear interpolation
between the nearest two grid frequencies and momenta is used. We can
use further symmetries of the translational invariant model to reduce
the computational effort. Besides time-reversal mentioned above, the
model features inversion and particle-hole symmetry. The latter symmetry
shuffles momenta (i.e. it moves the omitted $k_{2}$ momentum)
and thus the particle-hole transformed set of momenta generically is not on the grid. We thus combine this symmetry with complex
conjugation (and inversion symmetry) to find
\begin{eqnarray*}
U^{\Lambda}\left(\begin{array}{ccc}
\!k_{1}^{\prime} & k_{2}^{\prime}; & k_{1}\!\\
\!\Pi & X; & \Delta\!
\end{array}\right) & = & U^{\Lambda*}\!\left(\begin{array}{ccc}
\left|k_{1}^{\prime}+\pi\right| & \!\left|k_{2}^{\prime}+\pi\right|; & \!\left|k_{1}+\pi\right|\\
\Pi & X; & \Delta
\end{array}\right)
\end{eqnarray*}
compatible with the chosen grid points.
Together with inversion symmetry, the full, say $k_{1}$-dependence
of $U^{\Lambda}$ can be reproduced from the knowledge of its $k_{1}$-dependence
in one fourth of the BZ, say $k_{1}\in[-\pi,-\pi/2]$.

During the numerical integration of the flow equations for the model (\ref{eq:H}), we observe a divergence in the interaction vertex $U_{\Xi}\left(\omega_{min},0,\pi,0\right)$ (as well as in all symmetry-related terms) once the cutoff parameter $\Lambda$ becomes much smaller than the smallest Matsubara frequency $\omega_{min}$. This component describes scattering processes far away from the Fermi surface and we suspect the cause of the divergence
to lie in the van-Hove singularity of the density of states at the
points $k=0,\pi$. This is corroborated by the observation that in the absence of both interaction and self-energy feedback, the flow equation (\ref{eq:flow_kSpace_UXi}) evaluated at $\left(0;0,\pi,0\right)$ is governed by a term
\begin{equation}
 \textnormal{Re } \frac{1}{i\Lambda+2t\cos(q)}\frac{1}{i\Lambda+2t\cos(q+\pi)},
\end{equation}
which diverges for $\Lambda\to0$. Moreover, the flow is regularized by any terms which open a gap in the single-particle spectrum (e.g., dimerized hoppings). Since an exact solution of (\ref{eq:H}) does not feature any divergent scattering processes, their occurence in Eq. (\ref{eq:flow_kSpace_UXi}) is an artefact of the second-order fRG flow equations (similar artefacts appear in quantum dots with many degenerate levels \cite{Karrasch2007}); this is unsettling on general grounds. Pragmatically, we cut off the divergence by stopping the flow at the smallest positive Matsubara frequency, $\Lambda^{f}=\omega_{min}\ll t$. Since the divergent process is a high-energy one, this generally does not influence our results (see below for further discussions).

\paragraph{Application: Occupation numbers.}

The low energy theory of the homogeneous model (\ref{eq:H}) for $|U|\!<\!2t$
is a Luttinger liquid \cite{Giamarchi2003}. The fRG has a history
of studying various observables and correlation functions in microscopic
models for Luttinger liquids with impurities or disorder that lead
to power law behavior with respect to a spatial variable, see \cite{Andergassen2004,Metzner2012,Karrasch2015}.
The corresponding power law exponents are functions of the Luttinger
liquid parameter $K$ that is known from the Bethe-Ansatz solution, $K=\frac{\pi}{2\arccos\left(-U/2t\right)}$. It was shown that a fRG-1
scheme is capable to correctly reproduce the power law exponents to
first order in $U$. We now investigate if our fRG-2 scheme can reproduce
a power law with an exponent $\alpha$ of second order in $U$, $\alpha(U)=\mathcal{O}(U^{2})$.
Specifically, we investigate the zero-temperature occupation number
$n_{k}=\left\langle c_{k}^{\dagger}c_{k}\right\rangle $ for $k$
in the vicinity of the Fermi points which, as hallmark of Luttinger
liquid physics, for $k\rightarrow k_{F}$ asymptotically features
a power law \cite{Giamarchi2003}

\begin{equation}
\left|n_{k}-1/2\right|\propto\left|k-k_{F}\right|^{\alpha}\label{eq:nk power law}
\end{equation}
with
\begin{equation}
\alpha=\frac{K}{2}+\frac{1}{2K}-1.\label{eq:alpha}
\end{equation}
The exact $n_{k}$ for our microscopic model has been calculated with DMRG in Ref. \cite{Karrasch2012}. We take these data for comparison to our fRG-2 results.

\noindent 
\begin{figure}
\noindent \begin{centering}
\includegraphics{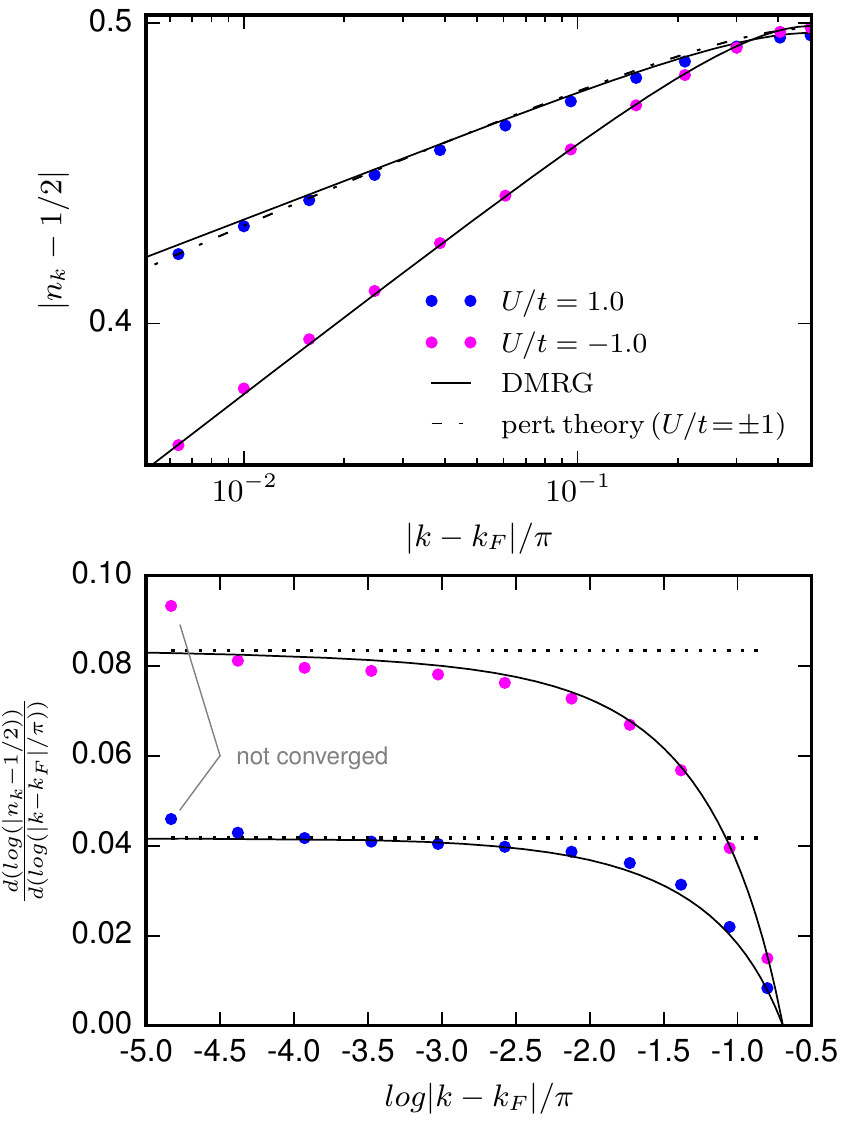}
\par\end{centering}
\caption{\label{fig:n(k)}The top panel shows fRG results for the occupation numbers
$n_{k}$ of a translationally-invariant system in the vicinity of the Fermi momentum for $U/t=\pm1$. The
solid lines show DMRG results from Ref. \cite{Karrasch2012} and the
dash-dotted lines denote the results of 2nd order perturbation theory (which coincide for $U/t=\pm1$).
To check if the fRG captures the power law nature of $n_{k}$, in
the bottom panel, we plot a log-log derivative of the data from the
top panel with asymptotic values of the Bethe ansatz exponent from Luttinger liquid
theory denoted by horizontal dashed lines that is also approached
by the fRG data.
}
\end{figure}

We compute the occupation number $n_{k}$ from the Green function
$G_{k}^{\Lambda_{f}}\left(\omega\right)$ analogous to Eq. (\ref{eq:occ}).
The results for $U/t=\pm1$ are shown in Fig. \ref{fig:n(k)}. In
the top panel, the fRG data (dots) show reasonable agreement with
the DMRG data from Ref. \cite{Karrasch2012} (solid lines) in the momentum
range considered. To answer the question if the asymptotic
power law form (\ref{eq:nk power law}) is captured in the fRG data,
we plot a discrete log-log-derivative in the bottom panel. Indeed,
the data in an intermediate k-range is compatible with the exponent
in Eq. (\ref{eq:alpha}) (dotted lines). In contrast, simple second
order perturbation theory (dash-dotted line) not only predicts the
same $n_{k}$ for $U/t=\pm1$ but asymptotically yields a logarithmic
behavior for $n_{k}$. Accordingly, this is signaled by a constant
simple log-derivative, which is not compatible the fRG-2
data (at least not for $U/t=-1$). The fact that fRG-2 calculations yield a power law in $n_{k}$ is consistent with the results of Ref. \cite{Honerkamp2003} where a simpler second-order scheme was used to analyze quasi-particle weights.

More solid evidence for power laws from fRG with correct exponents
would require $k$-data points closer to $k_{F}$. However, as can
be inferred from the data in Fig. \ref{fig:n(k)}, instabilities occur
for $|k-k_{F}|/\pi<10^{-2}$. These instabilities close to the Fermi
points are sensitive to almost all discretization parameters discussed
above. Despite an extensive exploration of parameter space and alternative
grid choices, no sense of convergence has been found. We suspect that
this has to do with the gapless nature of the free dispersion and
corresponding divergences in the Greens functions on the right hand
side of the flow equations in the limit $\Lambda\rightarrow0$ (c.f.
discussion above). Let us note that also changing to a momentum space
cutoff scheme does not cure the mentioned stability problems.

\section{\label{sec:Conclusion}Conclusion}

In conclusion, we have formulated a fRG scheme for spinless fermions
in one spatial dimension and thermal equilibrium that is correct up
to 2nd order in the interaction. We have considered both a real space
scheme for inhomogeneous systems and a momentum space scheme for homogeneous
systems. The latter allowed for a full feedback of spatial vertex dependence
in the flow equations and can be applied to any translationally-invariant model.
We have devised physical test scenarios to benchmark
both schemes in the realm of $T=0$ physics. On the upside, our fRG results compared
well with known analytic solutions and outperform both perturbation theory in general and existing
fRG schemes that are only correct up to first order in the interaction. On the downside, our fRG approach is correct \textit{only to second order}, and including higher terms seems futile. Moreover, the momentum-space approach suffers from unphysical instabilities in gapless systems.

For future work, we remark that the real space formalism can easily
be extended to include on-site potentials. Then the spinless fermion
model is a standard model to study the intriguing physics of many-body
localization. However, for this application, properties of excited
states have to be accessed. Another conceivable
future application would be the study of quantum phase diagrams. More
general types of interaction like a next-nearest-neighbor term $U^{\prime}$
can be readily included. Likewise, more general one-dimensional dispersions
can be studied with only minor modifications. In particular, interacting
symmetry protected topological phases are a promising candidate for
the application of fRG \cite{Sbierski}.

\section{Acknowledgment}

We acknowledge useful discussions with Jan von Delft, Christian Klöckner, Lisa Markhof,
Volker Meden and Jesko Sirker. Numerical computations were done on
the HPC cluster of Fachbereich Physik at FU Berlin. Financial support
was granted by the Deutsche Forschungsgemeinschaft through the Emmy
Noether program (KA 3360/2-1).

\bibliographystyle{apsrev}
\bibliography{/home/bjoern/Physics/PhD_FU/library}

\end{document}